\documentclass[aps,prd,twocolumn,superscriptaddress]{revtex4}
\usepackage{graphicx}
\usepackage{setspace}
\usepackage{multirow}

\mathchardef\mhyphen="2D
\def\dd{\ensuremath{d \mhyphen d}}

\begin{document}

\title{Measurement of the directional sensitivity of DMTPC detectors}



\author{Cosmin Deaconu}
\altaffiliation{Currently at the Kavli Institute for Cosmological Physics, University of Chicago, Chicago, Illinois 60637} 
\affiliation{Physics Department, and Laboratory for Nuclear Science, Massachusetts Institute of Technology, 77 Massachusetts Avenue, Cambridge, Massachusetts 02139}
\author{Michael Leyton}
\altaffiliation{Currently at Institute of High-Energy Physics, Barcelona Institute of Science and Technology, 08193 Bellaterra, Spain } 
\author{Ross Corliss} 
\affiliation{Physics Department, and Laboratory for Nuclear Science, Massachusetts Institute of Technology, 77 Massachusetts Avenue, Cambridge, Massachusetts 02139}
\author{Gabriela Druitt}
\affiliation{Royal Holloway University of London, Department of Physics, Egham, Surrey TW20 0EX, United Kingdom}
\author{Richard Eggleston}
\affiliation{Royal Holloway University of London, Department of Physics, Egham, Surrey TW20 0EX, United Kingdom}
\author{Natalia Guerrero}
\altaffiliation{Currently at MIT Kavli Institute, 77 Massachusetts Ave. Cambridge MA 02139 }
\affiliation{Royal Holloway University of London, Department of Physics, Egham, Surrey TW20 0EX, United Kingdom}
\author{Shawn Henderson}
\altaffiliation{Currently at Cornell University, Department of Physics, Ithaca, New York 14853 } 
\affiliation{Physics Department, and Laboratory for Nuclear Science, Massachusetts Institute of Technology, 77 Massachusetts Avenue, Cambridge, Massachusetts 02139}
\author{Jeremy Lopez}
\altaffiliation{Currently at Department of Physics, University of Colorado at Boulder, Boulder CO 80309}
\affiliation{Physics Department, and Laboratory for Nuclear Science, Massachusetts Institute of Technology, 77 Massachusetts Avenue, Cambridge, Massachusetts 02139}
\author{Jocelyn Monroe}
\affiliation{Royal Holloway University of London, Department of Physics, Egham, Surrey TW20 0EX, United Kingdom} 
\author{Peter Fisher}
\affiliation{Physics Department, Institute for Soldier Nanotechnology, MIT Kavli Institute and, Laboratory for Nuclear Science, Massachusetts Institute of Technology, 77 Massachusetts Avenue, Cambridge, Massachusetts 02139} 
\email[Corresponding author: ]{fisherp@mit.edu}
\date{\today}

\begin{abstract}

The Dark Matter Time Projection Chamber (DMTPC) is a direction-sensitive detector designed to measure the direction of recoiling $^{19}$F and $^{12}$C nuclei in low-pressure CF$_4$ gas using optical and charge readout systems. In this paper, we employ measurements from two DMTPC detectors, with operating pressures of 30--60\,torr, to develop and validate a model of the directional response and performance of such detectors as a function of recoil energy. Using our model as a benchmark, we formulate the necessary specifications for a scalable directional detector with sensitivity comparable to that of current-generation counting (non-directional) experiments, which measure only recoil energy. Assuming the performance of existing DMTPC detectors, as well as current limits on the spin-dependent WIMP-nucleus cross section, we find that a 10--20 kg scale direction-sensitive detector is capable of correlating the measured direction of nuclear recoils with the predicted direction of incident dark matter particles and providing decisive (3$\sigma$) confirmation that a candidate signal from a non-directional experiment was indeed induced by elastic scattering of dark matter particles off of target nuclei.

\end{abstract}

\pacs{29,95}

\maketitle

\section{Introduction}

The nature of dark matter presents the major challenge to the current theory of
particle interactions. Weakly Interacting Massive Particles (WIMPs), motivated
by supersymmetry and other theories with new physics at the 100 GeV energy
scale, provide an important candidate for dark matter. For thirty years, counting
experiments have sought detection of nuclear recoils induced by the elastic
scattering of neutral particles with 10--1,000~GeV/$c^2$ mass and $\beta\sim
0.001$, improving the cross section sensitivity from that of a very massive
Dirac neutrino, 10$^{-34}\,$cm$^2$, to the current limit of 10$^{-45}$\,cm$^2$. 

In the event of a statistically significant observation by a counting
experiment measuring only the recoil energy spectrum, confirmation that the
observed events resulted from the elastic scattering of dark matter particles
off of target nuclei will be crucial. While measurement with other target
isotopes may give some comfort that a candidate signal was caused by dark matter,
correlation with an astrophysical phenomenon will be essential. The motion of
the Solar System through the galactic dark matter halo provides two means of
establishing an astrophysical correlation: the annual modulation of the count
rate above a threshold energy and the sidereal variation in the recoil
direction of a struck target nucleus. Discussions surrounding the claimed
observations of annual modulation of the recoil rate have shown that this
method may be prone to instrumental and environmental
systematics~\cite{Cline:2015yza}; we have therefore pursued the more difficult, but
more decisive, sidereal directional modulation technique.

The recoil energy spectrum of nuclei struck by WIMPs falls exponentially with
energy with an $e$-folding factor proportional to the average WIMP kinetic
energy. The maximum nuclear recoil energy ranges from 5 to 250~keV, depending
on WIMP and target nucleus masses. Nuclear recoil experiments therefore place a
premium on low energy thresholds. A fluorine recoil with 40\,keV energy in
60\,torr of CF$_4$ gas will have a typical track length of $\mathcal{O}$(1\,mm).
Reconstructing the direction of such a recoil requires a detector with spatial
resolution of 300\,$\mu$m (or higher) to measure at least three points along
the track. At this pressure, probing meaningful cross-sections requires
detectors with tens or hundreds of cubic meters of target
volume~\cite{Ahlen:2009ev}.

We have carried out a performance study of a 20-liter DMTPC
detector~\cite{4sh}, scalable to a cubic meter, to understand whether loss of
directional information occurs due to physics processes, instrumentation, or
both. Our measurements support a model that allows us to assess the directional performance
of a cubic-meter DMTPC detector that we have built and
are currently commissioning~\cite{leyton_taup}. We show that an array of cubic
meter detectors could confirm or refute a claimed observation by the current
generation of counting experiments for spin-dependent interactions. (Due to the nuclear structure of $^{19}$F, DMTPC detectors are primarily sensitive to spin-dependent WIMP coupling.) We use our measurements to provide, for the
first time, a quantitative baseline for evaluating the detection technology of
direction-sensitive searches, and to identify places for improvement in the
directional technique. 

\section{Detector Description\label{se:desc}}

Time projection chambers (TPCs)~\cite{Fancher:1978es} achieve better than
100-$\mu$m spatial resolution in the drift direction over large sensitive
volumes. By using drift lengths of up to several meters to transport ionization
electrons from the site of a recoil event to an amplification and
readout plane, TPCs achieve high spatial resolution for large sensitive
volumes, at a low channel count. Proportional amplification gives
two-dimensional information on the recoil direction in the plane perpendicular
to the drift. The optical readout system in DMTPC images the amplification
plane and measures scintillation light produced during proportional
amplification, thereby measuring a two-dimensional projection of the recoil
onto the readout plane. Transient charge readout of the anode gives information
about the ionization distribution along the drift direction, i.e.\ the axis
normal to the amplification plane. Photomultiplier tubes (PMTs) measure
the total light output with nanosecond time resolution and give information
about the recoil along the drift coordinate direction.

In this paper, we model the performance of a 20-liter TPC with optical and charge readout systems, referred to as the 4Shooter \cite{4sh}. The cylindrical drift volume of the 4Shooter detector is housed within a set of field-shaping rings and measures 30.7\,cm in diameter and
26.7\,cm from cathode to anode, resulting in a sensitive volume of 19.8 liters of
CF$_4$ at 30--100~torr and a target mass of 2--10\,g. The ground mesh is 80\%
transparent and stands 435\,$\mu$m above the anode plane. For the measurements presented here, the anode was held at 670\,V, creating an electric field of 15\,kV/cm and a
measured gas gain of 67,000, calibrated using an $^{55}$Fe X-ray source. Typical drift fields were 180--200\,V/cm, chosen to minimize the diffusion of the electron swarm during the drift.

The optical system (4 $\times$ Canon 85~mm $f$/1.2 lenses, with a magnification
of 6.67, mounted onto 4 $\times$ Apogee Alta U6 CCD cameras with Kodak
KAF-1001E chips) has a geometric acceptance of 7$\times$10$^{-4}$, on average,
per camera and lens, for photons originating from the amplification region. The
four CCD cameras collect the scintillation light emitted during proportional
multiplication between the grounded mesh and anode plane. More details can be found in Ref.~\cite{4sh}. The cameras were operated in `witness' (continuous) mode, typically
imaging for one second before being read out. Transient charge and light
signals were collected during each exposure and stored along with the CCD
image. The optical system gain was calibrated using an $^{241}$Am $\alpha$ source, depositing approximately 4.0\,MeV per $\alpha$ in the sensitive volume and producing 10--19~counts/keV$_{ee}$, depending on the camera \cite{4sh}. Here, we use the subscript $_{ee}$ to denote electron-equivalent energy since not all of the recoil energy is converted into ionization, particularly for nuclei. Conversion factors between recoil energy and electron-equivalent energy are estimated using TRIM~\cite{srim}. Nuclear recoils were generated within the detector volume using AmBe and $^{252}$Cf neutron sources. 

For the measurements presented here, we also used a small chamber with a 10-cm-diameter `triple-mesh' amplification region, consisting of a shared anode mesh sandwiched between two ground meshes,
allowing optical readout of two back-to-back TPCs with a single camera. The gas
gain measured in this mode was about 100,000. We also operated the triple-mesh
amplification region in `cascade' mode, resulting in a maximum achievable gas
gain of approximately 10$^6$ at a gas pressure of 30\,torr for a single TPC.
However, most of our data with this chamber was collected with a gas gain of 440,000.
The optical system consisted of a Nikkor 55\,mm $f$/1.2 lens mounted onto an Andor Ikon
L936 camera. The optical system gain was estimated to be approximately
300\,counts/keV$_{ee}$. 

\section{Detector Response\label{sec:sim}}

\begin{figure*}
\centerline{\includegraphics[width=0.8\textwidth]{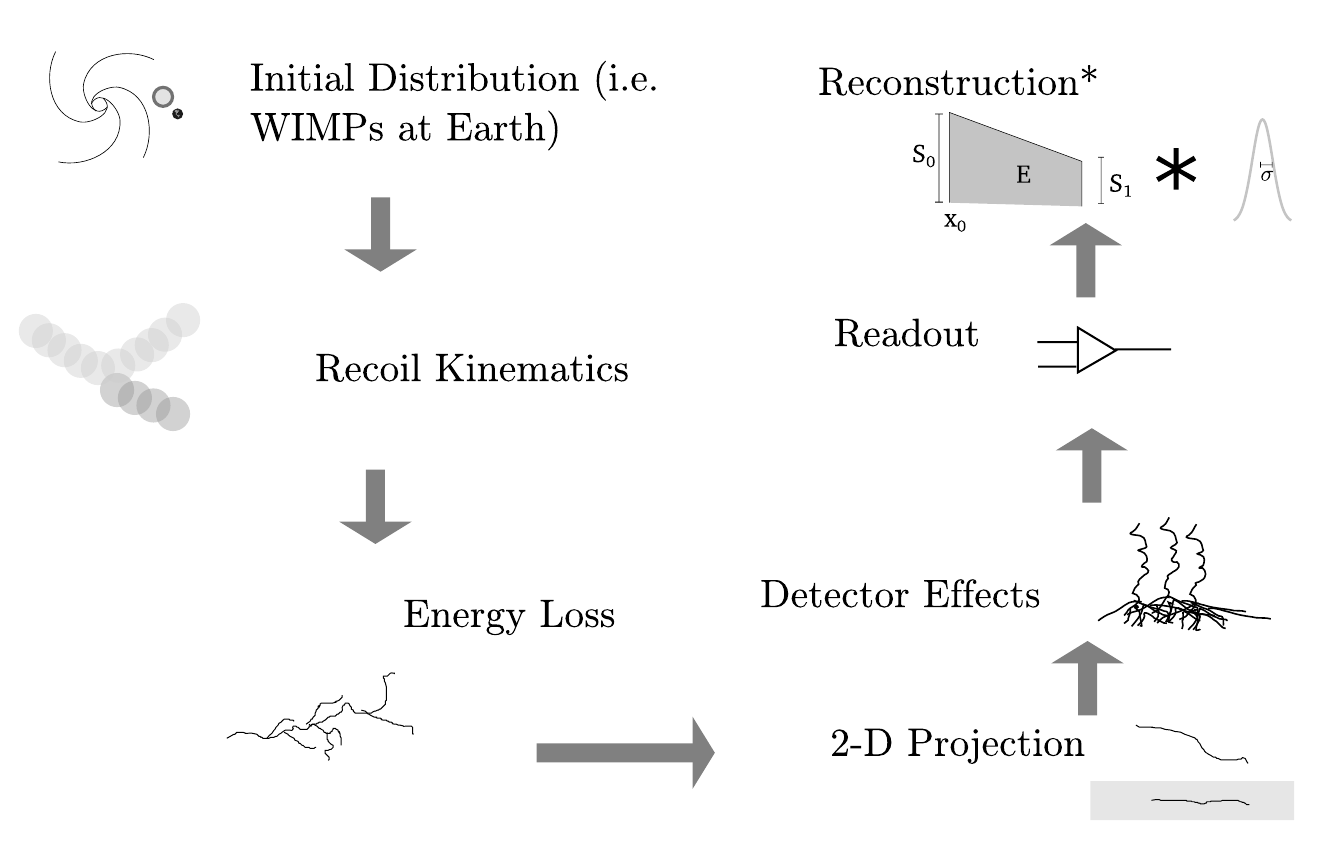}}
\caption{\label{fi:flow} Flow chart of events depicting the generation, amplification, detection and analysis of WIMP-induced elastic scattering. The $\ast$ indicates the fit function is a convolution of a linear energy loss with a two dimensional gaussian spatial resolution.}
\end{figure*}

We model the directional response of a DMTPC detector by simulating the steps shown
in Fig.~\ref{fi:flow} and comparing with calibration data collected from our
detectors. The sequence of events in the detector starts with the velocity
distribution of WIMPs near Earth and ends with the fit parameters of the
reconstructed track associated to a nuclear recoil induced by elastic
scattering of a WIMP with a $^{12}$C or $^{19}$F nucleus. In the study reported
here, the data input to the track fit is a CCD image of a nuclear recoil and
the output is the recoil direction in the amplification plane. This study does
not yet include information from the time structure of the charge readout in
the track reconstruction, which can also be used to determine the recoil angle
in the drift direction. Instead, information from the charge readout system has
been used to improve energy resolution and discriminate against backgrounds
coming from radioactivity of the internal components or cosmic rays passing
through the CCD sensors.

We simulate recoils of $^{19}$F or $^{12}$C nuclei due to incident WIMPs,
neutrons from a deuterium-deuterium (\dd) source and neutrons from an AmBe
source. For WIMP-induced recoils, we sample velocities from the Standard Halo
Model~\cite{shm} (SHM), which assumes an isotropic, isothermal sphere for the galactic dark matter distribution, and generate elastic recoils using two-body kinematics
with isotropic scattering in the rest frame. For AmBe and \dd\ sources, we
sample the neutron energy from the appropriate distribution, given the source
location outside of the detector. In both cases, we generate nuclear recoils
uniformly throughout the active volume of the detector.

\begin{figure}
\centerline{\includegraphics[width=\columnwidth]{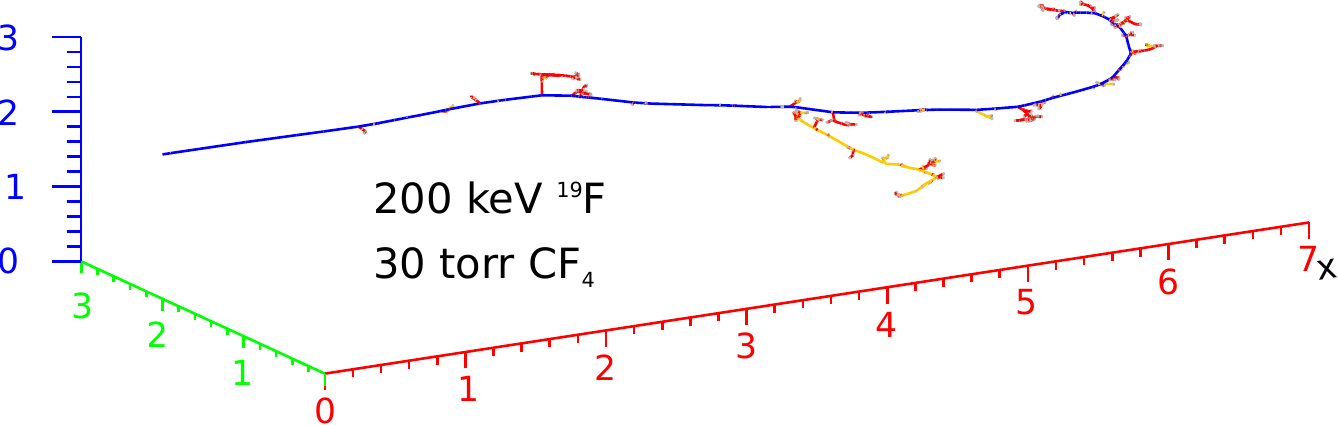}}

\caption{\label{fi:recoil}A TRIM-generated recoil cascade in 30\,torr CF$_4$.
The blue line represents the trajectory of the initial ion, a 200\,keV fluorine
recoil. The red and yellow lines represent the paths of secondary fluorine and
carbon nuclear recoils, respectively. The electron ionization is not shown. The
units on the plot are in mm.}

\end{figure}

Elastic scattering of WIMPs with masses in the range of 10--1,000~GeV/c$^{2}$ off of target nuclei with masses in the range of 10--20~amu impart up to 200\,keV of kinetic energy to the recoiling nucleus. Neutrons from AmBe and \dd\ sources induce nuclear recoils in the same range of energies. We simulate recoiling $^{19}$F or $^{12}$C nuclei with kinetic energies below 200\,keV. In this energy range, recoils lose energy via Coulomb interactions with atomic electrons (electronic stopping), which directly results in ionization, and via screened Coulomb interactions with atomic nuclei (nuclear stopping)~\cite{srim}. Nuclear stopping, which dominates over electronic stopping below approximately 50\,keV for $^{19}$F in CF$_4$, produces secondary ions that, then, also lose energy, resulting in indirect ionization losses by the primary ion. A single collision can produce energetic secondaries, causing the primary recoiling nucleus to scatter by a large angle, as illustrated in Fig.~\ref{fi:recoil}. We
use TRIM~\cite{srim} to simulate the secondary cascades from low-energy
$^{19}$F ions in low-pressure CF$_4$ in detail~\cite{cosminthesis}. The
trajectories of all recoils in the cascade in TRIM are then used to estimate the
three-dimensional ionization distribution resulting from the simulated primary recoil.

DMTPC measures the electrons liberated by the ionization of CF$_4$
molecules due to the motion of a recoiling $^{19}$F or $^{12}$C
nucleus. The work function of CF$_4$ is 34\,eV/pair \cite{wolfe}. An electric
field in the drift volume of $E = 190$\,V/cm transports the electrons towards
the amplification region with a velocity of 13\,cm/$\mu$s \cite{christophorou};
the field strength $E$ is chosen to minimize transverse diffusion. With the
4Shooter detector, we measured the ratio of the electron transverse diffusion
constant to the electron mobility, $D_T/\mu$, by parameterizing the transverse track width
$\sigma_{T}$ as a function of the drift distance, $z$:

\begin{eqnarray}
\sigma_{T}^2(z) = \sigma_{T,0}^2 + 2 \left( \frac{D_T}{\mu} \right) \left( \frac{z}{E} \right),
\end{eqnarray}

\noindent where $D_T/\mu = 0.053 \pm 0.005$\,V and $\sigma_{T,0} = 0.72 \pm 0.05$\,mm are
the best fit averages across cameras at a pressure of 60\,torr~\cite{4sh}. Our measured value for $D_T/\mu$ is
consistent with the literature, while the additional $\sigma_{T,0}$ term includes contributions from various effects such as the intrinsic track width, avalanche width, mesh grid
spacing, lens depth-of-focus, and camera resolution. The transverse diffusion is
approximately 1\,mm for a 25\,cm drift distance at 60\,torr.


Once at the ground plane, the ionization electrons are guided by the electric
field, through the 250-$\mu$m-pitch mesh, and into the amplification region. The
15\,kV/cm electric field causes proportional multiplication with a net
electron gain of up to 10$^6$ \cite{cosminthesis}. Scintillation photons are
produced during proportional multiplication 34\% of the time \cite{asher}. These photons image the electron swarm created by the nuclear recoil. We
calculate the electric potential in the amplification region due to the woven
mesh electrode structure using \texttt{gmsh}~\cite{gmsh} and
\texttt{ElmerFEM}~\cite{elmer}. The resulting potential map is then passed to a
\texttt{garfield++}~\cite{garfieldpp} library to perform the microscopic
simulation of the avalanche, recording the spatial distribution of the
ionization. For the 4Shooter detector, simulation suggests that the avalanche adds
100\,$\mu$m to the transverse width of the track.  Production of scintillation
light is simulated by sampling the ionization distribution and transporting the
scintillation photons through the optical viewport and lens to the CCD camera.
This step takes view factors and light attenuation of the optics into account.
Simulation of the camera response to the incident photons includes the
scintillation wavelength spectrum and the CCD quantum efficiency, as well as
the measured camera bias and read noise.

\subsection{Readout and Reconstruction\label{se:fit}}

Ref.~\cite{cosminthesis} describes the offline processing of the CCD images and
simulated recoils in detail. A brief summary is presented here. In the case of
simulated recoils, the camera bias level and read noise from data are added to
the simulated images. Images are then cleaned to remove CCD artifacts such as
hot pixels, cosmic rays and residual bulk images. Next, dark frames are used to
subtract pedestal offsets between pixels and the optical system gain
calibration from the $^{241}$Am source is applied. Track finding begins by
low-pass filtering the image to improve the signal-to-noise ratio for pattern
recognition, followed by a custom hysteresis-thresholding segmentation
algorithm~\cite{cosminthesis} to build clusters around seed pixels with counts
above threshold. Neighboring clusters are merged, particularly when separated
by known dead regions of the detector. The clusters are then cleaned of pixels
below a minimum threshold.  The resulting clusters correspond to the
two-dimensional projections of the electron swarms onto the amplification
plane.  A final classification step identifies the cluster as a spark, residual
bulk image, CCD artifact, cosmic ray, $\alpha$ track, or a nuclear recoil
inside or outside of the fiducial region.  Only the last category of events is
used for this directional study.  Non-nuclear-recoil events are removed by
applying the same set of cuts in data and simulation, described in detail in
Ref.~\cite{cosminthesis}. 

Track parameters from selected clusters associated to nuclear recoils are
estimated in the following way for both data and simulation. The intensity
values of the pixels comprising the track are modeled as $I(x,y) =
G(x,y)S(x,y)+ N(x,y)$, where $x$ and $y$ refer to the position on the CCD chip,
$G$ is the spatially-dependent system gain (counts/ionization) and $S$ is the
best fit of the track ionization density model, averaged over each pixel. $N$
is the predicted number of noise counts in each pixel and is modeled as a
combination of Gaussian camera read noise and Poisson shot noise. $S(x,y)$ is
the convolution of a Gaussian model of the diffusion and avalanche spreading in
the amplification region with a line segment with linearly varying ionization
density. We find that this line segment model is a useful approximation to the
Bragg curve. Seven parameters fully characterize the track:  ionization energy
$(E_I$), one end of the track ($x_0$,$y_0$), the track axis ($\phi$), the
initial ionization density ($S_0$), the change in ionization density over the
length of the track ($\Delta S$), and the convolution width ($\sigma$).
\texttt{Minuit2}~\cite{minuit} carries out the minimization, with initial
values based on a principal component analysis of the intensity-weighted pixels
belonging to a track.

The angle $\phi$ gives the reconstructed average axis of ionization of the
recoil in the amplification plane and $\Delta S$ provides the direction, or
sense, along the axis defined by $\phi$. For recoil energies below the Bragg
peak ($\sim$1\,MeV for fluorine), the ionization profile ($dE/dx$) decreases
with energy. The asymmetry in ionization density along the track direction is
used as an estimator of the vector direction. Determining the sign of $\Delta
S$ presents the key challenge of this work, referred to as the sense or
``head-tail'' assignment of the track along the axis defined by $\phi$. For
nuclear recoils below $\sim 1$\,MeV, $\Delta S<0$ means the ion lost more
energy at the start of the track than at the end and that
$\left(x_o,y_o\right)$ refers to the start of the track, while $\Delta S>0$
means that $\left(x_o,y_o\right)$ refers to the end of the track.

\begin{figure}
\includegraphics[width=\columnwidth]{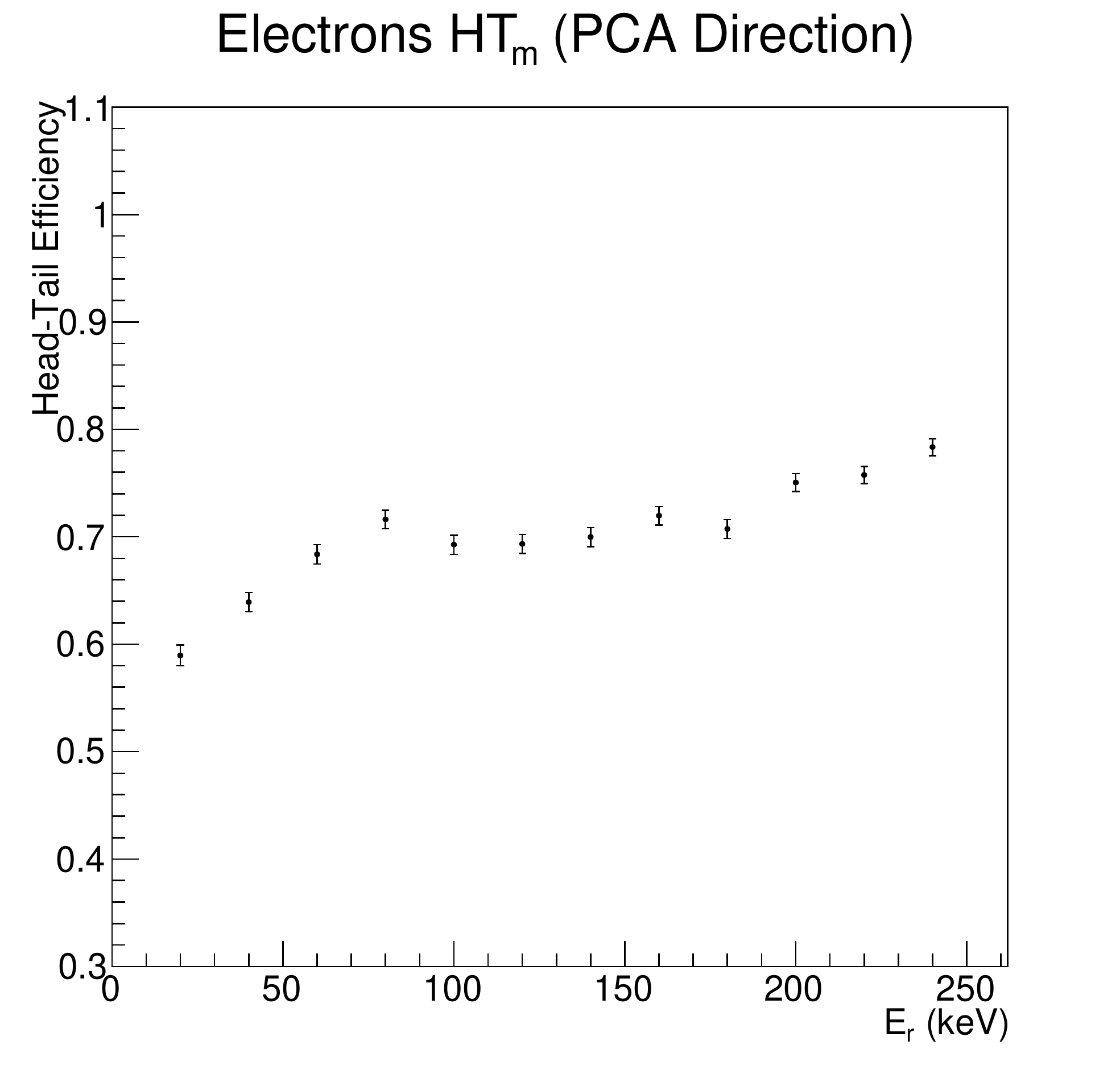}
\caption{The simulated fraction of recoils at 30\,torr assigned the correct vector sense, or ``head-tail'', based on the slope of a line fit, $\Delta S$, to the ionization density deposited onto the CCD chip using a principal component axis, prior to any detector effects. }
\label{fig:intrinsic_ht} 
\end{figure}

Fig.~\ref{fig:intrinsic_ht} shows the probability, or efficiency, of correctly
assigning the head-tail sense in simulated recoils without consideration of any
detector effects. A value of 1.0 on the ordinate of Fig.~\ref{fig:intrinsic_ht}
means that the correct head-tail assignment is always made. Guessing blindly
corresponds to a value of 0.5, meaning the guess is correct half of the time.
This plot shows that there is considerable loss of information before any
detector effects are considered, coming mainly from nuclear collisions during
the stopping of the primary recoiling ion in the surrounding gas. We will
return to this point later, as this is an important point for the detector
performance: the fraction of recoils assigned the correct sense in Fig. 3 shows
the maximum possible efficiency. A perfect detector using this direction
assignment method would measure a head-tail fraction of 0.7 at a recoil energy
of 100\,keV.

\section{Experimental Measurement of Directional Performance}

We have carried out three measurements aimed at quantifying the directional performance of DMTPC detectors, namely how well they measure the axis and sense of a recoiling nucleus. This section describes each measurement and compares it with predictions from the simulation.
 
 \subsection{Measurements using $\alpha$ particles} 

The directional performance is studied using measurements of $\alpha$ particles with known position and direction. We simulate low-energy recoil nuclei
by placing a collimated $^{241}$Am source above the detector cathode of the
4Shooter detector such that only the last few hundred keV of the $\alpha$'s
enter the fiducial volume. This configuration generates low-energy $^4$He tracks
at a shallow angle at the maximum drift distance from the anode. While the
directional response to low-energy $^4$He is not interesting for dark matter
searches in pure CF$_4$, the simplicity of the setup allows for a
well-controlled test of the simulation model. For this measurement the 4Shooter detector
was operated at 60~torr, with a gas gain of 67,000.  

We measure the gain using $^{55}$Fe and $^{241}$Am sources mounted inside the vacuum vessel. $^{55}$Fe emits photons with energies of 5.9 keV and 6.5 keV that produce electrons in CF$_4$ gas via photoelectric absorption. $^{55}$Am emits x-rays with energies of 13.9, 17.5, and 21.1 keV \cite{jeremythesis}.  We compared the gas gain measured with Cremat CR-112  and CR-113 charge integrating amplifiers with  gains of 13 and 1.3 mV/pC, respectively, and found agreement at the 2\% level. Using a work function of 34 eV/pair for CF$_4$ and carrying out the gain calibration at several pressures gives an additional uncertainty of 5\%. Combining data from both $^{55}$Fe and $^{241}$Am sources gives a calibration linear to within 1.5 \cite{jeremythesis}. Quenching factors from TRIM \cite{srim} are also included, giving results comparable to those measured by the MIMAC collaboration  \cite{mimacquenching}, who reported a quenching factor of 0.38 in CF$_4$ at 50 mbar (37.5 torr) for $^{19}$F recoils with 20 keV of energy.

\begin{figure*}
\includegraphics[width=0.8\textwidth]{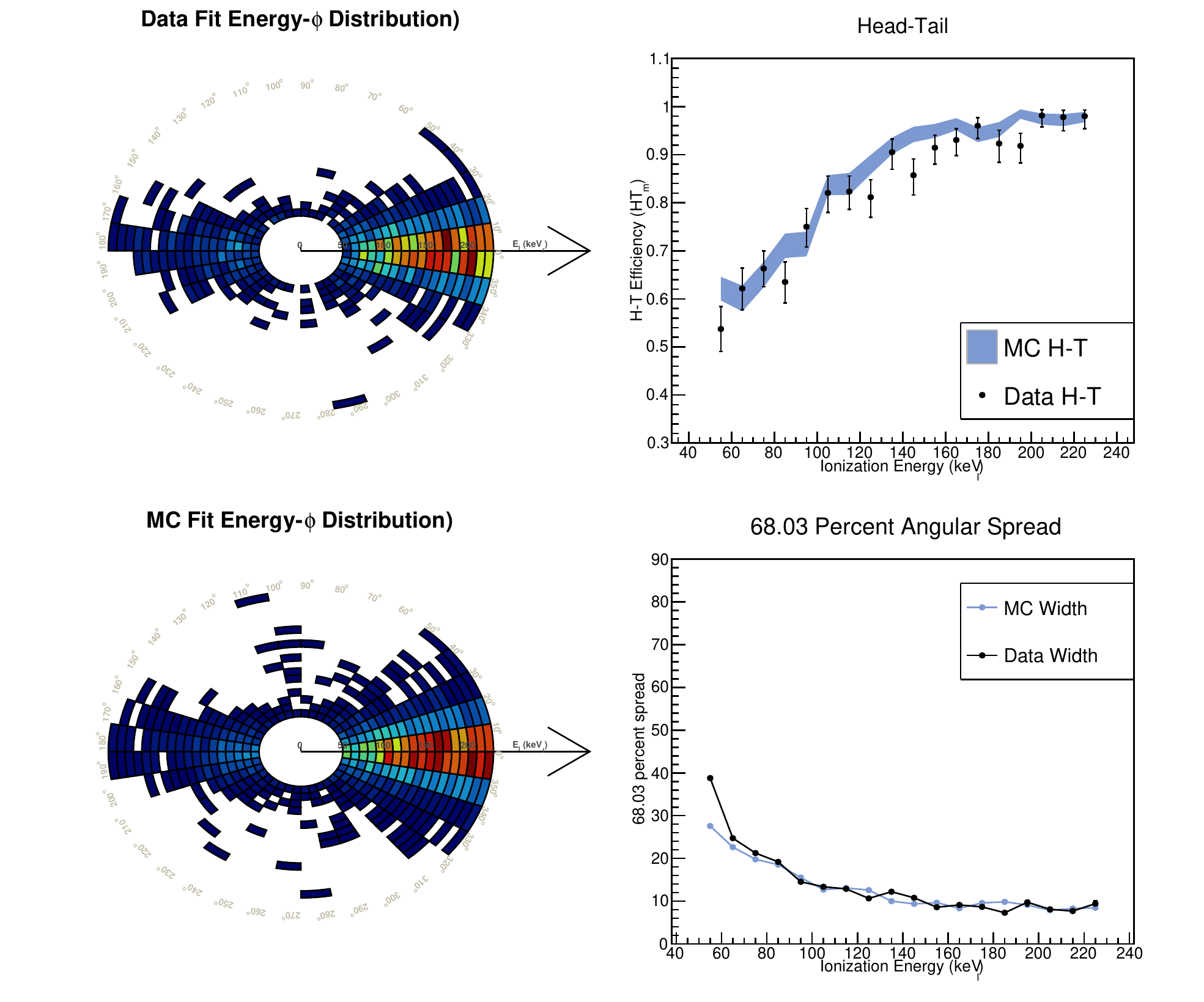}
\caption{The directional response to angled $\alpha$'s for data and simulation. The head-tail efficiency plot shows the fraction of tracks reconstructed with the correct sense, while the axial spread plot shows the angular spread (in $^{\circ}$) containing 68\% of tracks. The colors on the energy-$\phi$ plots (left) scale linearly from $\sim 0$ (blue) to a maximum of 0.02 (red).}
\label{fig:aa}
\end{figure*}

The full detector
simulation, adjusted to match the measured system gain, was used to simulate the
same scenario. The directional response of both data and simulation are shown
in Fig.~\ref{fig:aa}, indicating generally good agreement between the two.
Simulation predicts a level of head-tail assignment a few percent better than
we find in the data, while the angular spread of recoils is at a similar level as predicted. This gives confidence in the simulation model of the gas physics
and recoil response.

\subsection{Measurements using high-energy neutrons\label{se:fourshooterperf}} 

A neutron source is used to study the directional performance with nuclear recoil tracks, similar to a WIMP-induced signal. We used a Troxler Laboratories 3320 AmBe fast neutron source positioned near the 4Shooter detector to produce low-energy $^{19}$F and $^{12}$C recoils. The detector collected data
for 5.4 live days. The source was located several meters from the detector, sufficiently far to ensure a collimated beam of incoming neutrons, but sufficiently close to sustain a relatively high neutron flux. 
The gain calibration described in the previous section is used for this study. 

The AmBe source produces neutrons through an $\alpha \mhyphen n$ process, where
approximately $10^{-4}$ of the $\alpha$'s from the $^{241}$Am decay produce
a neutron via $\alpha\, +\,^{9}\mathrm{Be} \, \to \, ^{13}\mathrm{C}^* \to
\, ^{12}\mathrm{C} \, + \, n$. The resulting neutron energy spectrum has
several peaks and extends up to approximately 12\,MeV. Ref.~\cite{iso} provides
a reference spectrum, but the actual spectrum depends on the details of the
construction of the source. A two-inch lead-brick shield was placed in front of the neutron beam in order to reduce the rate of sparking in the amplification region induced by the high rate of $\gamma$ rays from the $^{237}$Np decay. We used a \texttt{Geant4}-based~\cite{geant4} simulation to account for the
neutron interactions in the lead brick. The resulting neutron energy spectrum
is broadly similar to that of Ref.~\cite{iso}, but the simulated spectrum has a
larger average recoil angle with respect to the source direction. We simulate
events between 40 and 200\,keV$_{ee}$ and find that, with a modest track
reconstruction fit quality requirement of $\chi^2$/$n_{\mathrm{dof}} < 2$, the
efficiency of reconstructing nuclear recoil tracks in this energy range with
the 4Shooter detector is 36\%.

\begin{figure*}
\includegraphics[width=\textwidth]{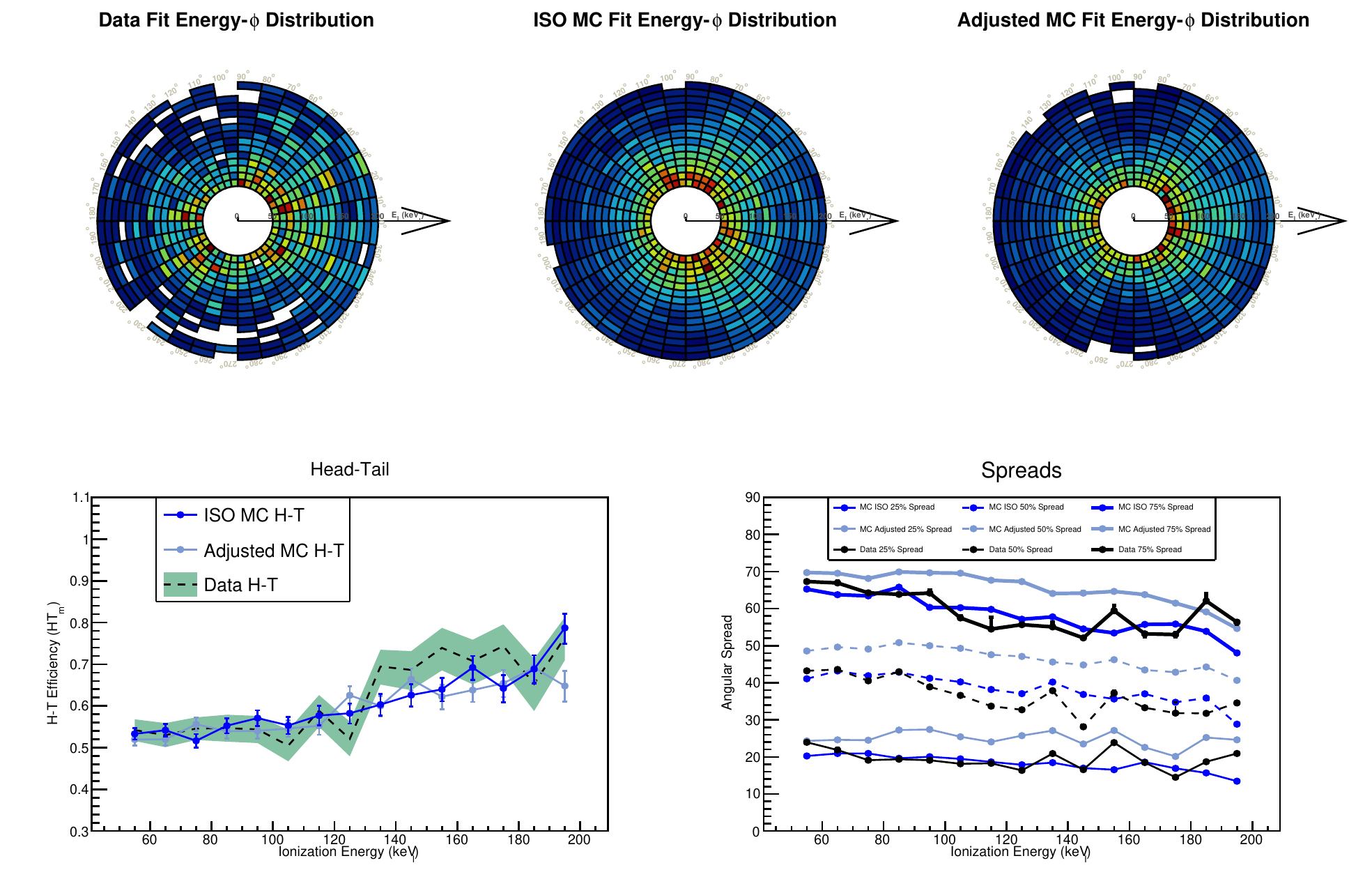}
\caption{Measured and simulated energy-angle spectra from the AmBe source. Two different assumptions about the initial neutron spectrum are used, with and without lead shielding, as described in the text. The colors on the energy-$\phi$ plots (top) scale linearly from $\sim 0$ (blue) to a maximum of 0.006 (red).}
\label{fig:ambe}
\end{figure*}

Fig.~\ref{fig:ambe} shows the energy-angle spectrum measured in data and
predicted by the simulation. The angular spread shows 25, 50, and 75\%
quantiles, since due to kinematics, the peak of the recoil direction spectrum
is not expected to be in the mean direction of the neutrons at low energies.
The simulated head-tail efficiency is generally in agreement with that measured
in data, while the predicted angular spread in direction is 10--20\% larger in
simulation. This is likely due to uncertainties in the angular distribution of
neutrons emitted by the source. 

\subsection{Measurements with low-energy neutrons} 

We studied the directional performance versus lower recoil energies by illuminating the 10-cm test chamber, described in Section~\ref{se:desc}, with a deuterium-deuterium (\dd) neutron generator designed by Schlumberger.  A \dd\
neutron generator fuses deuterons via the reaction $d \, + \, d \to \,^3\mathrm{He} \, +
\, n$, which produces neutrons with an energy of 2.45\,MeV and results in nuclear recoils with
a maximum possible recoil energy of $\mathcal{O}(500)$\,keV.  X-ray sources are used to determine the gas gain, as described above. The location of the Bragg peak at 21 keV/mm for CF$_{4}$ at 30 torr \cite{cosminthesis} serves as a useful cross check with the gain calibration. The general agreement between data and MC for the energy-range recoil distributions shows that the energy scale is linear to within 10\% across the range 25 -- 500 keV.

In this detector, the nuclear recoil detection efficiency is estimated to be 35\% for tracks in the range of 5--200~keV$_{ee}$ by comparing data and simulation, with the same recoil event selection and reconstruction quality cuts as described above. For this study, the high gas gain of the cascaded amplification system caused a small non-Gaussian effect, described in the Appendix in more detail. We accounted for this effect by adding a second Gaussian to the track fit described in Section~\ref{se:fit} and fitting for the two additional parameters.

\begin{figure*} 
  \includegraphics[width=\textwidth]{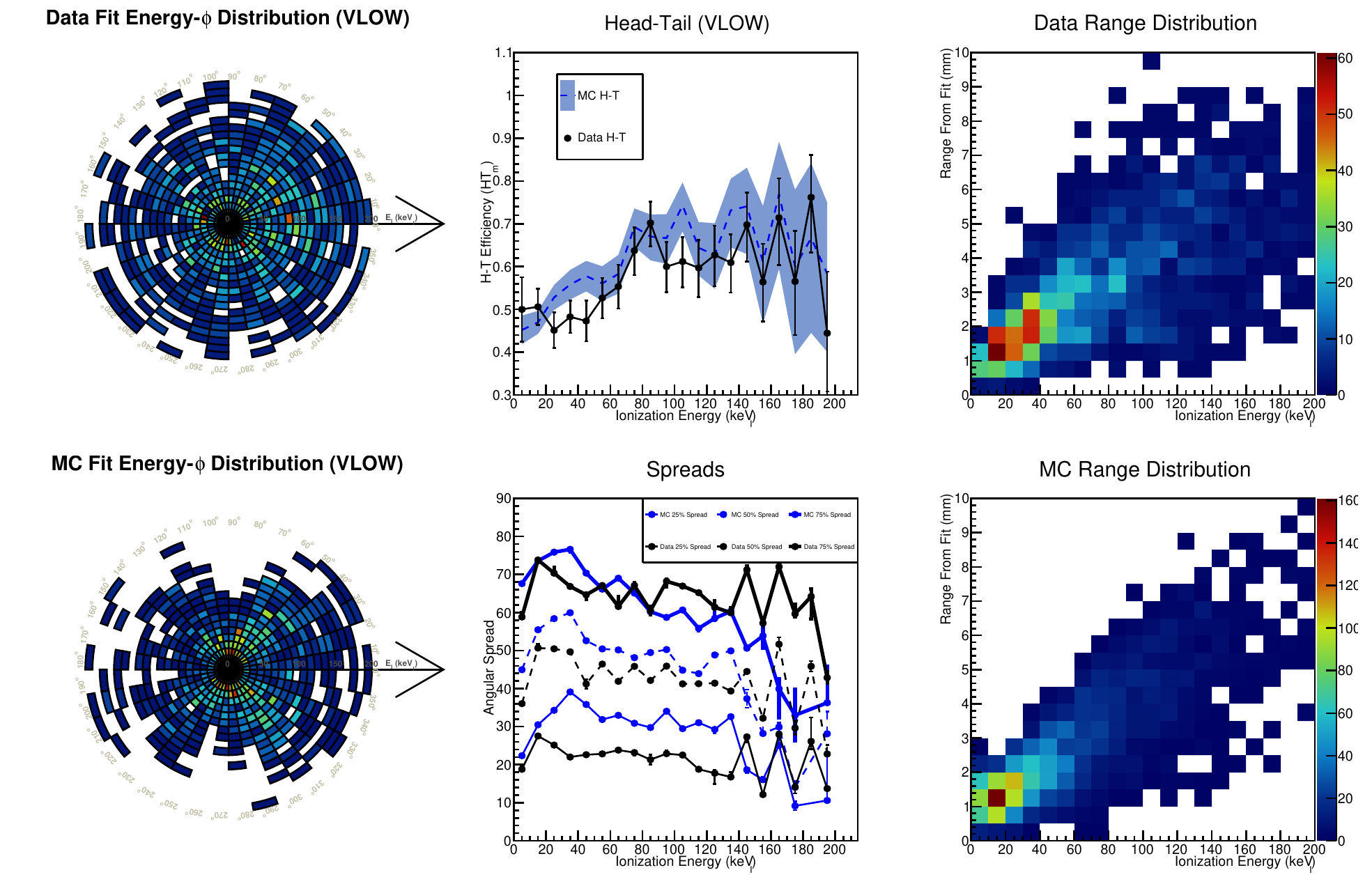}

\caption{Comparison of the directional response between data and simulation to
a \dd\ neutron generator using a test stand equipped with a cascaded
triple-mesh amplification region. The simulation models the neutrons as
monochromatic, although approximately 10\% are expected to interact in the
generator casing or chamber wall, possibly explaining the difference in axial
response. The range-energy distributions (right) between the data and
simulation are comparable. The colors on the energy-$\phi$ plots (left) scale
linearly from $\sim 0$ (blue) to a maximum of 0.008 (red).} 

\label{fig:dd} 
\end{figure*}

Fig.~\ref{fig:dd} compares the directional response of the detector between
data and simulation and shows that the reconstructed range-energy distributions are broadly consistent between the two. Similar to the previous study with AmBe neutrons, the angular spread of the recoil
directions predicted by the simulation is larger than that observed in the data, whereas the data contain a
larger fraction of recoils pointing along the mean direction of the incident neutrons.  A likely explanation is that the simulation assumes mono-energetic neutrons, while we calculate that roughly 10\% of the neutrons interact in the generator casing or chamber wall before reaching the fiducial volume, which
modifies the energy spectrum of incident neutrons. Additional modeling of the
neutron propagation may be able  to produce better agreement.

Fig.~\ref{fig:dd} shows good agreement between the head-tail assignment efficiency in simulation and in data, which approaches 70\% for recoil energies above 140\,keV. By comparison, the maximum measurable head-tail efficiency
before any detector effects, shown in Fig.~\ref{fig:intrinsic_ht}, is also 70\%
at a recoil energy of 140\,keV. This is an important benchmark, demonstrating
that the DMTPC detector technology discussed here successfully measures the
intrinsic directionality of the recoil signal, and that the fundamental physics
limit of this approach is the straggling of the primary $^{19}$F or $^{12}$C
ion in the target gas. This property of CF$_4$ as a target gas is also
relevant to other target gases, such as CS$_2$, and therefore applicable to all
current TPC-based directional experiments.

\section{Directional Sensitivity}

In this section, we develop a metric for quantifying the directional
performance of DMTPC detectors using the axial and head-tail measurements of
recoiling nuclei discussed in the previous section. We then use this metric to
outline the specifications of a directional detector capable of establishing
whether a putative signal from a current-generation non-directional, counting
experiment has a sidereal variation in direction.

Given the large fluctuations in energy loss at low energies, the performance
metric defined here combines both the axial direction reconstruction and the
head-tail assignment to utilize all available directional measurement
information. We define an opening angle, called the axial spread, as the angle
containing a specified fraction of tracks originating about the incident source
direction. Using the directional response simulation, validated with data as
described above, we find that a better measure of the performance results from
combining the axial spread with the head-tail assignment. 

We construct the directionality metric as follows:  suppose a background-free
detector observes $N$ candidate WIMP events, each with a reconstructed
ionization energy and direction. For the set of recoils generated by those $N$
interactions, we separate the directional response into head-tail and axial
components and bin these variables in recoil energy, into bins of width $\Delta E_R$. We
compute the reconstructed forward fraction with respect to the expected WIMP
direction, $HT(E_R)$, and the axial spread, $W(E_R)$. For a given energy bin of
$HT(E_R)$ and $W(E_R)$, we calculate the probability that the observed value (or
larger) could have arisen from an isotropic background distribution. We combine
the $p$-values for each bin using Fisher's method~\cite{fishermethod} to create
an overall isotropy rejection statistic.

For an isotropic background distribution, axial angles with respect to the
expected WIMP axis are uniform between zero and ninety degrees. For the
$i^{\mathrm{th}}$ energy bin containing a sample of $m$ events, of which $k$
events are along the expected WIMP axis, the probability of observing $HT>k/m$
is

\begin{eqnarray}
p\left(HT>k/m\right) &=& I_{1/2}\left(k+1/2,m-k+1/2 \right),
\label{eq:ht}
\end{eqnarray}

\noindent where $I_{1/2}$ is the regularized incomplete beta
function~\cite{eadie}, which is a continuum version of the binomial distribution.  The subscript 1/2 gives the probability of forward vs. backward scattering and the additional factors of 1/2 in Eq.~\ref{eq:ht} are included in order to improve the coverage for a discrete distribution.
For an isotropic background, the probability that half
the tracks fall in a wedge of opening angle $\phi<W\left(E_r\right)$ around the
expected WIMP axis is

\begin{equation}
  p\left(\phi\leq W\right) =  I_{W/90^{\circ}}\left( \lfloor m/2 \rfloor +1, 
  \lfloor m/2 \rfloor +2 \right),
  \label{eqn:p_w}
\end{equation} 

\noindent where $\lfloor m/2 \rfloor$ is the floor of $m/2$. The subscript $W/90^{\circ}$ in Eq.~\ref{eqn:p_w} gives the probability for an event from an isotopic distribution for fall in an angle $W$ around the WIMP axis. We compute the head-tail and
axial probabilities $p_i$, as in Eqns.~\ref{eq:ht} and \ref{eqn:p_w}, for each
energy bin, giving $2s$ degrees of freedom, where $s$ is the total number of
energy bins. We then combine the probabilities $p_i$ into a $\chi^2$ statistic,

\begin{eqnarray}
\chi^2_{2s} &=& -2 \sum_{i=0}^{2s} \log p_i.
\label{eq:chi2_f}
\end{eqnarray}

We calculate the Fisherian $p$-value for rejection sensitivity, $p_r$, which is
the probability that a measurement arises from the null (in this case isotropic
distribution) hypothesis. This is the CDF of the $\chi^2_{2s}$ distribution,

\begin{eqnarray}
p_r&=&\frac{\gamma\left(s,-\sum_{i=0}^{2s} \log p_i \right)}{\Gamma\left(s\right)}.
\label{eq:chi2}
\end{eqnarray}

\noindent where $\gamma\left(s,x\right)$ is the incomplete gamma function and
$\Gamma\left(s\right)$ is the gamma function. In this way, we combine the
head-tail statistic $HT$ and spread statistic $W$ for all energy bins. 

We have verified that the resulting test statistic provides approximately uniform coverage for an isotropic input (i.e. a value of $p_r$ = 1\% occurs about 1\% of the time), making it a valid metric for rejecting isotropy \cite{cosminthesis}. We do not claim to have developed the optimal test statistic, but instead focus on the main result of the paper, which is the performance of the metric on simulated WIMP recoil data, validated by the measurements described above. It is possible that a refined statistic could provide greater rejection power.

We now have the tools needed to quantify how well a directional detector can
measure a directional signal for a given WIMP mass and rejection level $p_r$.
We model a directional detector with a fiducial volume of one cubic meter and
reconstruction performance as described in Section~\ref{se:fourshooterperf},
operating at a pressure of 30~torr and a gas gain of 100,000.

We start by generating 100 pseudo-experiments, each with $N$
dark-matter-induced recoils ranging from $N=50$ to 1,000 in increments of 50
events. The WIMP velocities are drawn from the three-dimensional
Maxwell-Boltzmann distribution of the Standard Halo Model. We simulate two-body elastic scattering of
WIMPs with mass $M_{\chi} = 10$, 30, 100, 300 and 1,000\,GeV/$c^{2}$ off of
$^{12}$C or $^{19}$F, with recoil kinetic energies above 25\,keV, which is the
approximate simulated track-detection threshold in the CCD. We model the
full detector response and reconstruct the energy, axial direction, and
head-tail assignment of each recoil track as described in
Section~\ref{sec:sim}. Finally, we compute $p_r$ from Eq.~\ref{eq:chi2} for the
ensemble of $N$ recoils in each pseudo-experiment.

\begin{figure*}
  \centering
  \includegraphics[width=0.8\textwidth]{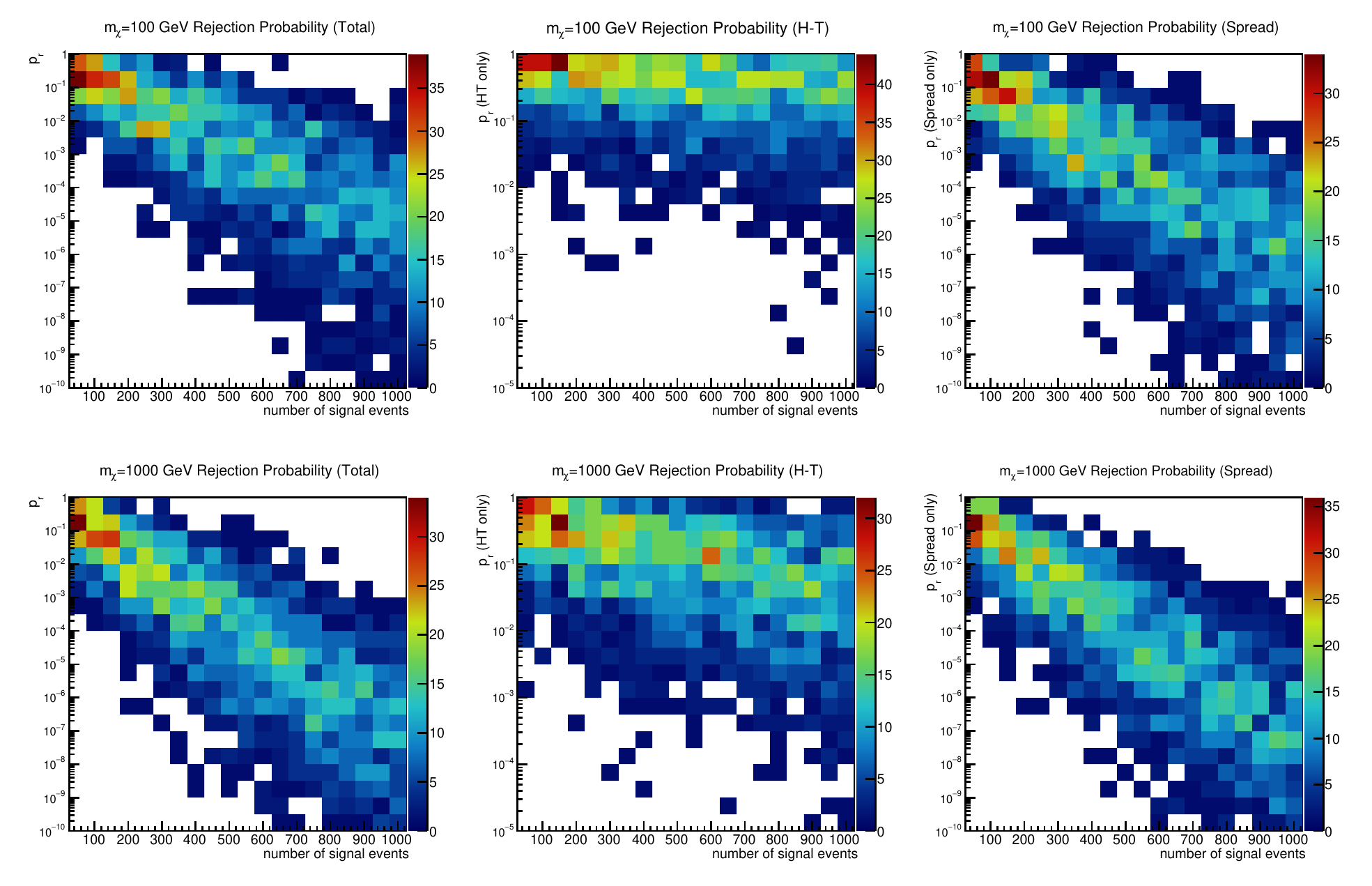}
  
  \caption{Isotropy rejection ($p_r$) as a function of the number of signal
    events, $N$, for WIMPs with mass $M_{\chi} = 100$\,GeV/$c^{2}$ (top) and 1000\,GeV/$c^{2}$(bottom). The leftmost
    column shows the total rejection, the center column shows rejection from
    sense (head-tail) only, and the right column shows the contribution to
    rejection from the axial measurement. The color scale shows the percentage of pseudo-experiments for a fixed number of signal events.
  }
  \label{fig:example_rejection}
\end{figure*}

Fig.~\ref{fig:example_rejection} shows the isotropy rejection for WIMPs with
$M_{\chi} = 100$ and 1000\,GeV/$c^{2}$ , using $HT$ or $W$ only, and $HT$ and
$W$ combined. For $M_{\chi} = 100$\,GeV/$c^{2}$, $HT$ provides little rejection
power, owing to limited intrinsic head-tail efficiency at low energies. $HT$
becomes more powerful for $M_{\chi} = 1000$\,GeV/$c^{2}$ since the recoil
energy spectrum is harder.

\begin{figure} 
  \centering
  \includegraphics[width=\columnwidth]{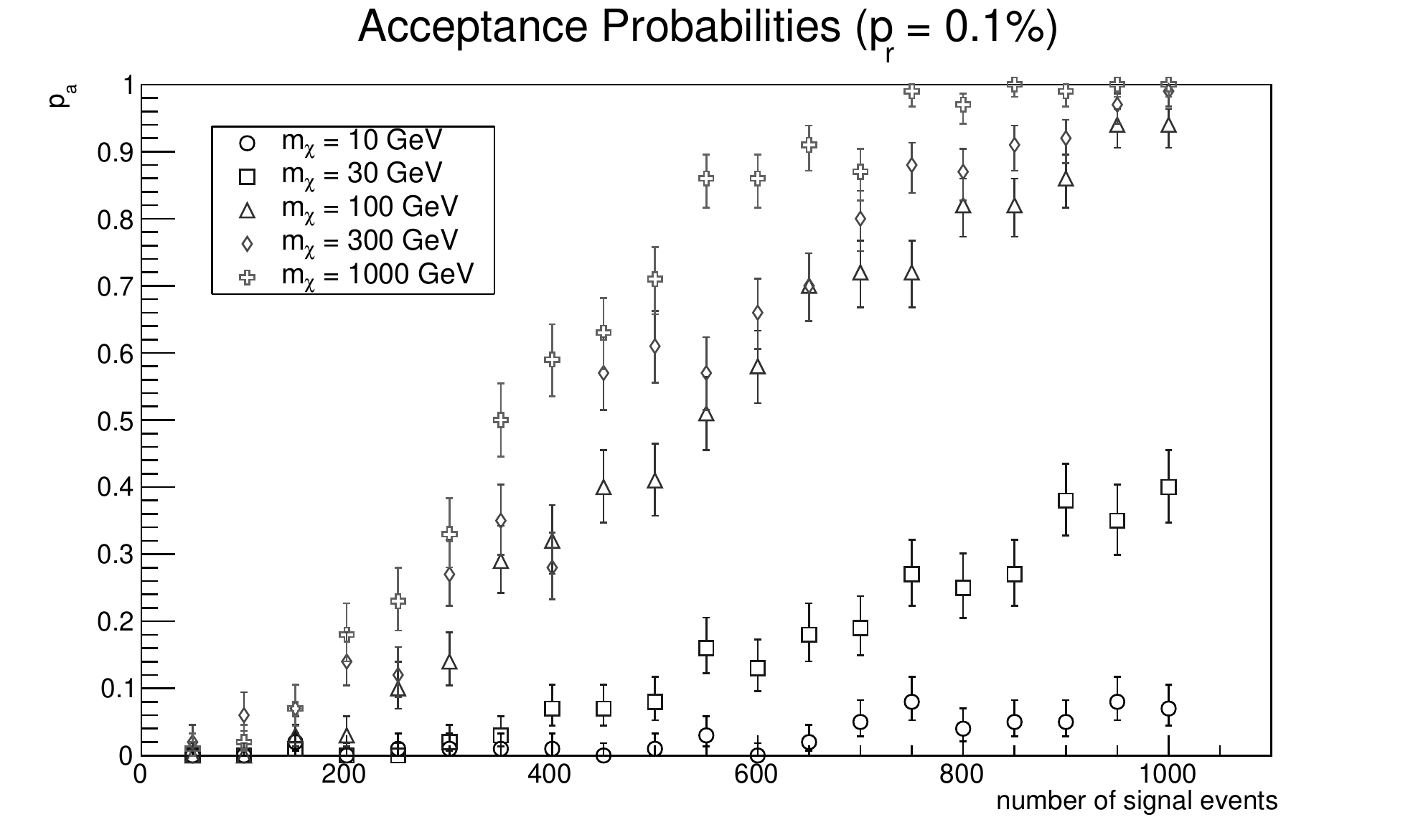}
  \caption{The fraction of pseudoexperiments $p_a$ achieving $p_r = 0.001$ for a given number of signal events $N$, for various simulated WIMP masses.}
  \label{fig:acceptance} 
\end{figure}

From the pseudo-experiments at each energy, we estimate the acceptance
probability $p_a(N)$, or the fraction of experiments achieving rejection
probability $p_r$ for a given number of signal events. We require $p_r = 0.001$, corresponding to 3$\sigma$ rejection. The results for all simulated WIMP masses are
shown in Fig.~\ref{fig:acceptance}. For $M_{\chi} = 100$~(300)\,GeV/$c^{2}$, 550~(450) events are
required for rejection at the 3$\sigma$ level in approximately half of the
pseudo-experiments. 

The number of required events can be reduced by selecting only those with
reasonable directional reconstruction confidence. A head-tail assignment
quality metric is derived from the fit used in reconstruction: after the
initial fit, the fit is repeated forcing the opposite sense $\Delta S$, and the
likelihood ratio of the two senses is used to derive a head-tail quality
metric. By cutting on the head-tail confidence such that the upper 50\% of
events are selected, we find that the number of required events to establish
3$\sigma$ rejection of isotropy is reduced by 23\% (17\%) to 425 (375) total
events (before the selection cut is applied) for $M_{\chi} =
100$~(300)\,GeV/$c^2$. 

We can now calculate the exposure (target mass $\times$ live time) required for
a detector to measure the number of events above a given energy threshold
needed to reject isotropy at the level of $p_r=0.1\%$. For this, we combine formulations from Ref.~\cite{shm} (Eq.\ 3.9) and \cite{McCabe} (Eq.\ 1) to calculate the differential rate, $dR/dE_{R}$, of dark matter signal events as a function of recoil energy, $E_{R}$:

\begin{eqnarray}
\label{eqn:differential_rate}
\frac{dR}{dE_{R}} & = & \frac{R_0}{E_0 r} \frac{1}{2 \pi v_{0}^{2}} \int_{v_\mathrm{min}}^{\infty} \frac{d^{3}v}{v} f(\vec{v} + \vec{v}_E)
\end{eqnarray}

\noindent where 
\begin{eqnarray}
  R_0 & = &\frac{490.43}{M_\chi M_T} \left(\frac{\sigma_0}{1\mathrm{\,pb}}\right) \left(\frac{\rho_D}{0.39\mathrm{\,GeV\,cm^{-3}}}\right)\\ &&  \nonumber \left(\frac{v_0}{230\mathrm{\,km\,s^{-1}}}\right) 
\mathrm{(kg \mhyphen day)}^{-1},
\end{eqnarray}
\begin{eqnarray}
E_0 & = & \frac{1}{2} M_\chi v_{0}^{2},
\end{eqnarray}
\begin{eqnarray}
r & = & \frac{4 M_\chi M_T}{(M_\chi + M_T)^2},
\end{eqnarray}
\begin{eqnarray}
\frac{v_\mathrm{min}}{c} & = & \frac{M_\chi + M_T}{M_\chi} \sqrt{\frac{E_R}{2 M_T}},
\end{eqnarray}

\noindent $M_\chi$ is the WIMP mass, $M_T = 0.932\,A$\,GeV/$c^{2}$ is the target mass, $\sigma_0$ is the WIMP-nucleus cross section for zero momentum transfer, $\rho_D$ is the local dark mater density and $\vec{v}_E$ is the Earth velocity relative to the dark matter distribution.  $f(\vec{v})$ is the Maxwell-Boltzmann velocity distribution in the galactic frame, truncated at the galaxy escape velocity, $v_{\mathrm{esc}}$, and $v_0$ is the dispersion velocity. An analytical expression for the integral in Eq.~\ref{eqn:differential_rate} is given in Appendix B of Ref.~\cite{McCabe}. We use $\rho_D = 0.39$\,GeV/cm$^{3}$~\cite{Catena}, $| \vec{v}_E | = 244$\,km/s, $v_{\mathrm{esc}} = 544$\,km/s, and $v_0=230$\,km/s here. To account for suppression of the cross section at large momentum transfer, we additionally include the spin-dependent form factor from Eq.~4.5 of Ref.~\cite{shm}. 

Table~\ref{ta:sum} shows the number of cubic-meter-detector days required to
detect one signal event for various WIMP masses, given a spin-dependent
WIMP-fluorine cross section $\sigma_{0,F} = 1$\,pb or a spin-dependent
WIMP-proton cross section $\sigma_{0,p} = 1$\,fb. The equivalent values are listed for a spin-dependent WIMP-proton cross section $\sigma_{0,p} = 0.49$\,fb, corresponding to the 95\% upper limit predicted by a Constrained Minimal Supersymmetric Standard Model (CMSSM)~\cite{CMSSM} for $\mu > 0$, where $\mu$ is the Higgs/higgsino mass parameter. The detector performance
demonstrated with the 4Shooter is assumed here, with an operating pressure of
30\,torr of CF$_4$ gas and a fluorine target mass of 120~g. For WIMPs with mass
100~(300)\,GeV/$c^{2}$ and $\sigma_{0,F} = 1$\,pb, there will be one signal
event, on average, every 62 (154) live days in a cubic-meter detector at the
specified conditions and performance. Note that the rows corresponding to
$\sigma_{0,p}$ require making standard spin-dependent
assumptions~\cite{Tovey2000} and include various spin factors, as well as the reduced mass of the WIMP-proton collision system.

\begin{table*}
\def\arraystretch{1.3}
\begin{tabular}{ l | c c c c c }
\hline
\hline
\multirow{2}{*}{$M_{\chi}$ (GeV/$c^2$)} & \multirow{2}{*}{~~~~~~10~~~~~~} & \multirow{2}{*}{~~~30~~~} & \multirow{2}{*}{~~~100~~~} & \multirow{2}{*}{~~~300~~~} & \multirow{2}{*}{~~1,000~~}\\
& \\
\hline
Percentage of recoils & \multirow{2}{*}{0.038} & \multirow{2}{*}{15.5} & \multirow{2}{*}{37.8} & \multirow{2}{*}{45.9} & \multirow{2}{*}{48.8} \\
above 25\,keV \\
\hline
Exposure per event (m$^3$-days) & \multirow{2}{*}{6,678} & \multirow{2}{*}{46} & \multirow{2}{*}{62} & \multirow{2}{*}{154} & \multirow{2}{*}{481} \\
for $\sigma_{0,F}$ = 1\,pb \\
\hline
Exposure per event (m$^3$-days)~~~~~ & \multirow{2}{*}{147,063} & \multirow{2}{*}{364} & \multirow{2}{*}{272} & \multirow{2}{*}{541} & \multirow{2}{*}{1,563} \\
for $\sigma_{0,p}$ = 1\,fb \\
\hline
Exposure per event (m$^3$-days)~~~~~ & \multirow{2}{*}{300,129} & \multirow{2}{*}{743} & \multirow{2}{*}{554} & \multirow{2}{*}{1,105} & \multirow{2}{*}{3,191} \\
for $\sigma_{0,p}$ = 0.49\,fb \\
\hline
\hline
\end{tabular}
\caption{\label{ta:sum} Expected exposures (in cubic-meter-detector days) for various dark matter masses, given a spin-dependent WIMP-fluorine cross section ($\sigma_{0,F}$) of 1\,pb or a spin-dependent WIMP-proton cross section ($\sigma_{0,p}$) of 1\,fb or 0.49\,fb~\cite{CMSSM}. An operating pressure of 30\,torr of CF$_{4}$ gas and a fluorine target mass of 120\,g have been assumed here.}
\end{table*}

\begin{figure}
  \centering
  \includegraphics[width=\columnwidth]{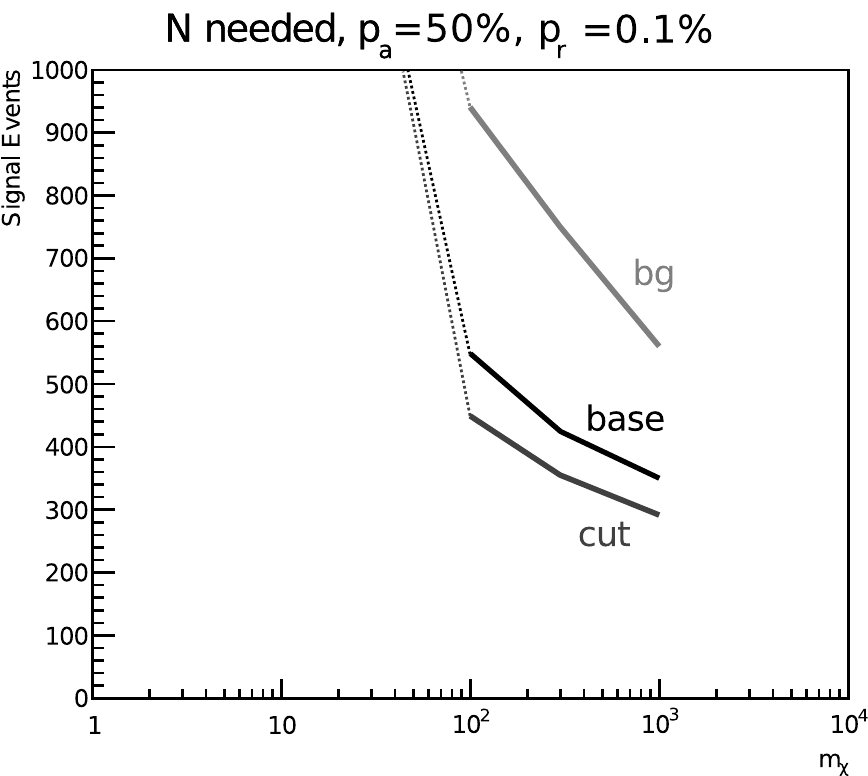}
  \caption{Number of signal events required to reject the isotropic hypothesis
at 3$\sigma$ for 50\% of pseudo-experiments, as a function of WIMP mass ($M_{\chi}$), given a WIMP-fluorine cross section $\sigma_{0,F} = 1$\,pb. The curve labeled ``base'' corresponds to Fig.~\ref{fig:acceptance}, while the ``cut'' and ``bg'' curves show the effect of applying a quality cut on head-tail reconstruction or adding an equal number of isotropic background events as signal events, respectively. The number of events for the ``cut'' curve corresponds to the number of signal events before any cut is applied.}
  \label{fig:variations} 
\end{figure}

The main result of this study is presented in Fig.~\ref{fig:variations}, which
shows the number of events needed to reject the isotropic hypothesis at
3$\sigma$ for 50\% of pseudo-experiments, as a function of WIMP mass, given a
WIMP-fluorine cross section of 1\,pb. This result accounts for the full
directional response of the detector, from straggling of the primary ion,
through reconstruction of the recoil track axis and head-tail assignment. For a
WIMP mass of 100 (300)~GeV/$c^2$, 550 (450) events are needed to reject
isotropy at the 3$\sigma$ level half of the time. If a quality cut on the
head-tail assignment is applied, only 425 (375) total events are needed, before
selection. Using Table~\ref{ta:sum}, the latter case with head-tail quality cut
translates to an exposure of 26,400 (57,800) cubic-meter-detector days. This
exposure is equivalent to an array of approximately 70 (160) cubic-meter
detectors, or a single cubic detector with a fiducial length of 4.2 (5.4)\,m,
operating for one year at 100\% live time. Assuming a pressure of 30\,torr
CF$_{4}$, this corresponds to a fluorine target mass of 8.7 (19)\,kg. If the
WIMP-proton cross section is 1\,fb, the same array of detectors would require
4.4 (3.5) years at 100\% live time to achieve the same sensitivity. This sets
the scale for the experiment, using the current measured DMTPC detector
performance.

The analogous statement for the current spin-independent cross section limits
of 10$^{-45}$\,cm$^2$ requires 10$^{11}$ cubic-meter-detector days, probably
outside the limits of any low-pressure gas target detector. 

Note that this discussion assumes perfect background rejection, i.e.\ that all
recoils measured have been induced by WIMPs.  While DMTPC has demonstrated excellent electron recoil rejection across a broad energy range \cite{DmtpcBackground}, nuclear recoils from fast neutrons are indistinguishable on an event-by-event basis and will have approximately the same energy spectrum as that of elastic WIMP scattering, although these studies were carried out at a higher threshold. To study the effect of background on the sensitivity, an equal number of isotropically distributed nuclear recoil background events is
added to the signal events, with the same energy spectrum as the signal. The
result is shown in Fig.~\ref{fig:variations}. As expected, the sensitivity is
degraded by the presence of backgrounds: for a WIMP mass of 100
(300)\,GeV/$c^2$, 925 (750) events are needed to achieve the same sensitivity. 

\section{Outlook}

This work has estimated the number of events required to reject isotropy in the
distribution of candidate dark matter events using a full model of experimentally measured detector
directional response for the first time. The model has been validated by
detailed comparison with data of the reconstructed axial angle and head-tail
assignment. Such a measurement would provide decisive evidence that a candidate
dark matter signal is associated with the dark matter halo of our galaxy.

Improvements in sensitivity beyond the DMTPC detector performance presented
here require improved head-tail efficiency at lower recoil energies. The model
introduced here may be used to evaluate different detector configurations. If
the projected 2-D electrons at generator-level `truth' are used to estimate the
sense, rather than the corresponding reconstructed track, only 81 events are
required to achieve the same sensitivity for a WIMP mass of 100\,GeV/$c^{2}$.
This gives an indication of the fundamental physics limit from ion straggling
and shows that an alternative detector configuration, with e.g.\ a different
gas target or medium, could provide up to a factor of five better sensitivity.
Further improvements beyond this level would require targets with lower nuclear
stopping at low energies, in order to reduce straggling of the primary ion and
preserve more information about sense in the ionization distribution. Potential
targets with lower mass sensitive to spin-dependent interactions are H and
$^3$He. The softer resultant recoil spectrum would likely require a higher gas
gain in the amplification region, perhaps using the triple mesh in the cascaded
configuration.

Alternatively, optimizing for axial direction reconstruction at
increased operating pressure may be a sound strategy in light of the
increasingly low limits on dark matter interaction cross sections.
The model introduced here can be used to study the trade-offs between
axial reconstruction performance, head-tail sensitivity, target type,
and target mass.

In summary, for spin-dependent WIMP interactions, an array of 70--160
cubic-meter DMTPC detectors, or a single cubic fiducial volume measuring 4.2--5.4\,m on one side (assuming that the effects from diffusion can be controlled), could make a decisive (3$\sigma$) determination at $\sigma_{0,p} \sim 1$\,fb half of the time, with an exposure of approximately 4 live years, for WIMP masses between 100 and 300\,GeV/$c^{2}$, assuming no background. There may be a factor
of five improvement in performance with better targets and detector readout,
but the energy straggling of the primary ion associated with the nuclear stopping power presents a significant barrier to further improvements in TPCs using gases such as CF$_4$ or CS$_2$.

 \appendix

\section{High-gain avalanches}

\begin{figure}
\centerline{\includegraphics[width=\columnwidth]{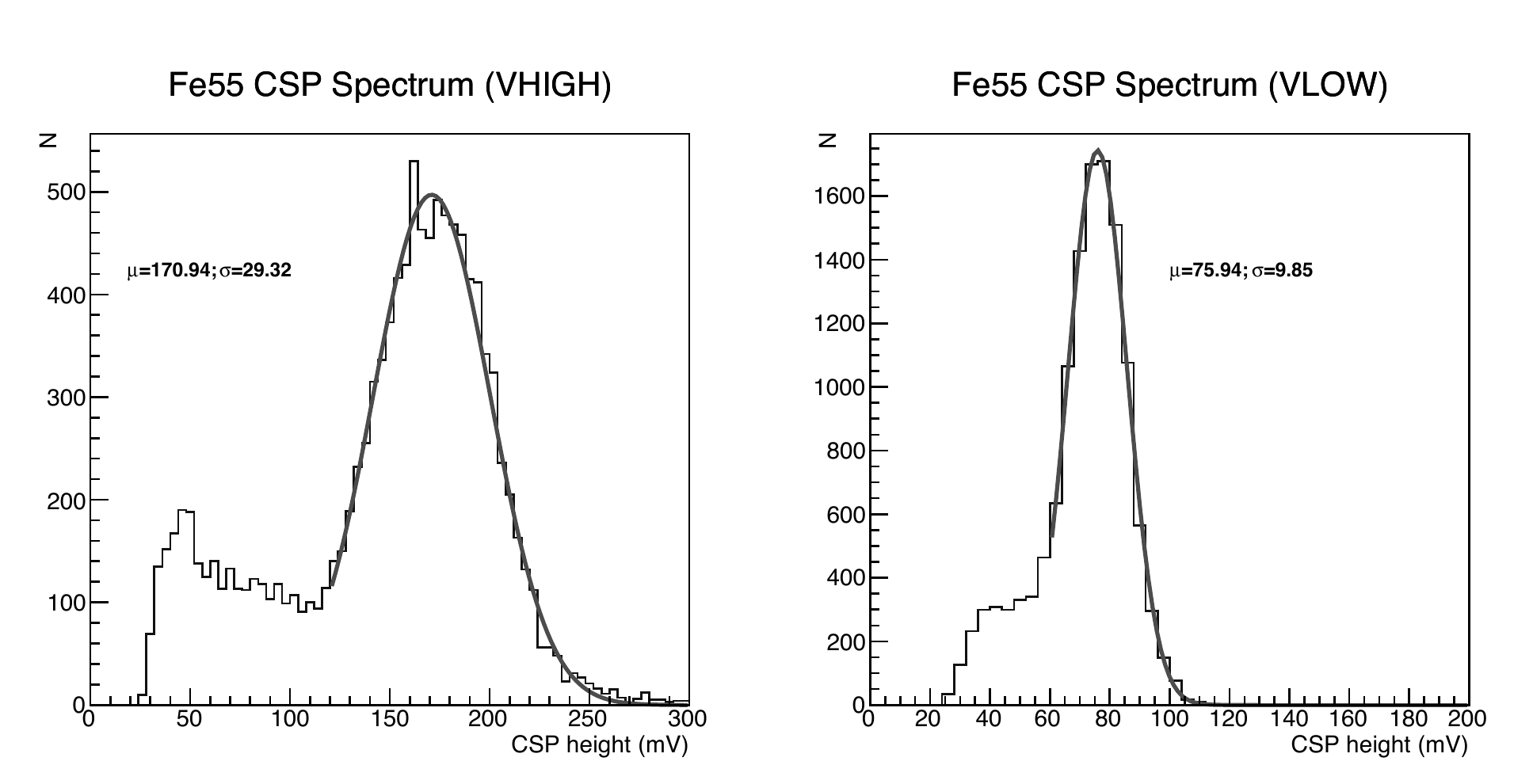}}
\caption{\label{fi:higain}Charge-sensitive preamplifier spectra of the 10 cm chamber with cascaded amplification regions at two different voltage settings with an $^{55}$Fe source inside. Based on the 5.9 keV expected peak energy and preamplifier gain, the inferred gas gains are 437,000 and 984,000.}
\end{figure}

Our 10 cm chamber is used to investigate operation at gains above 10$^5$, Fig.~\ref{fi:higain}.  
In the 10-cm chamber operating at a gain above 10$^5$, we observe a peculiar
feature of tracks associated to nuclear recoils: the tails of their transverse
projection are non-Gaussian. Our simulation does not reproduce this feature.
One explanation involves rare electron-impact processes producing states which
can decay into ionizing ultraviolet photons. The UV photons travel up to 1\,mm
in the gas, larger than the avalanche size, before ionizing, providing a
mechanism for non-Gaussian track widths. Measurements in
Ref.~\cite{ionizingvuv} indicate that there are processes at electron kinetic
energies of 200\,eV that produce UV photons. Inserting these rare processes
into the simulation qualitatively produces long tails, but also results in a
much higher gain since the photons travel in the direction opposite the
electric field, then ionize, creating a new avalanche. If this is indeed the
mechanism responsible, there must also be some quenching that is not included
in our model; \texttt{garfield++} does not include space charge effects, which
could provide an explanation. Since the simulation does not reproduce the gain
and spatial distribution simultaneously, the simulation was performed without
the ionizing photons. The analysis of the data is therefore adjusted for the
effective non-Gaussian convolution kernel by convolving with a sum of two
Gaussians and adding two parameters (the second Gaussian width and ratio of
amplitudes) to the fit. This shape matches the data well and could be motivated
by the presence of two independent mean-free-paths (electron and UV photon). 

\begin{acknowledgments} 
  
This work was supported by NSF PHY-1004592, DOE DE-SC0011970, STFC
ST/K502261/1, ERC-2011-StG 279980, and the Royal Holloway University
of London Crosslands Scholarship. We additionally thank Igal Jaegle
for providing the $\texttt{Geant4}$ simulation of the ISO spectrum
moderated by a lead brick. Alex Leder provided valuable assistance
with the \dd\ neutron generator.

\end{acknowledgments}

\bibliography{DirectionalityPaper}

\end{document}